\documentclass[conference]{IEEEtran}
\usepackage[utf8]{inputenc}
\usepackage[T1]{fontenc}

\usepackage{graphicx}
\usepackage{bmpsize}
\usepackage{caption}
\usepackage{subcaption}

\usepackage[dvipsnames]{xcolor}

\usepackage{lipsum}
\usepackage{blindtext}

\usepackage{hhline}
\usepackage{multirow}
\usepackage{tabularx}
\usepackage{booktabs}

\usepackage{floatrow}
\newfloatcommand{capbtabbox}{table}[][\FBwidth]

\usepackage{diagbox}

\usepackage{makecell}

\usepackage{footmisc}

\usepackage{amssymb}
\usepackage{amsthm}
\usepackage{csquotes}
\usepackage{mathtools}
\usepackage[linesnumbered, lined, noend, algoruled]{algorithm2e}
\usepackage{amsmath}
\usepackage{stmaryrd}
\usepackage{semantic}
\usepackage{tabularx}
\newcolumntype{C}[0]{>{\centering\arraybackslash}X}
\usepackage{arydshln}
\usepackage{siunitx}
\usepackage{amsmath}

\usepackage{fancybox}
\usepackage{tikz}
\usetikzlibrary{arrows.meta}
\definecolor{xred}{HTML}{DA4531}
\definecolor{xgreen}{HTML}{64AA37}
\definecolor{xblue}{HTML}{3131DA}
\usepackage{wrapfig}
\usepackage[all]{xy}
\usepackage[curve]{xypic}
\usepackage{tikz-inet}

\newif\ifcomments
\commentstrue 

\definecolor{anthonycolor}{HTML}{AA0000}
\definecolor{nicolascolor}{HTML}{00AA00} 
\definecolor{xaviercolor}{HTML}{0000AA} 
\definecolor{bhargavcolor}{HTML}{DB7093} 

\definecolor{blue}{HTML}{0000FF}
\definecolor{red}{HTML}{FF0000}
\definecolor{green}{HTML}{00FF00}

\newif\ifoldsota
\oldsotafalse 

\ifcomments
\newcommand{\anthony}[1]{\textcolor{anthonycolor}{[Anthony: #1]}}
\newcommand{\nicolas}[1]{\textcolor{nicolascolor}{[Nicolas: #1]}}
\newcommand{\xavier}[1]{\textcolor{xaviercolor}{[Xavier: #1]}}

\else
\newcommand{\anthony}[1]{}
\newcommand{\nicolas}[1]{}
\newcommand{\xavier}[1]{}
\fi

\newcommand{\g}[1]{\textbf{#1}}

\usepackage[numbers]{natbib}

\ifCLASSINFOpdf
\else
\fi
\usepackage{todonotes}
\usepackage{csquotes}

\begin{document}
%

\title{A Simple Reservoir Model of Working Memory with Real Values}

\author{\IEEEauthorblockN{Anthony Strock$^{2,1,3}$, Nicolas P. Rougier$^{1,2,3}$ and Xavier Hinaut$^{1,2,3,*}$}
\IEEEauthorblockA{1- INRIA Bordeaux Sud-Ouest, Talence, France\\
2- LaBRI, Université de Bordeaux, CNRS UMR 5800, Talence, France\\
3- IMN, Université de Bordeaux, CNRS UMR 5293, Bordeaux, France\\
$*$ Corresponding author}}


%


\maketitle

\begin{abstract}
The prefrontal cortex is known to be involved in many high-level cognitive functions, in particular, working memory. Here, we study to what extent a group of randomly connected units (namely an Echo State Network, ESN) can store and maintain (as output) an arbitrary real value from a streamed input, i.e. can act as a sustained working memory unit. Furthermore, we explore to what extent such an architecture can take advantage of the stored value in order to produce non-linear computations. Comparison between different architectures (with and without feedback, with and without a working memory unit) shows that an explicit memory improves the performances.
\end{abstract}

\begin{IEEEkeywords}
Working Memory, Gated Memory, Reservoir Computing, Prefrontal Cortex, Echo State Networks, ESN
\end{IEEEkeywords}

%
\IEEEpeerreviewmaketitle

\section{Introduction}

Prefrontal cortex (PFC), noteworthy for its highly recurrent connections~\cite{goldman1987circuitry}, is involved in many high level capabilities, such as decision making \cite{bechara1998dissociation}, working memory \cite{goldman1987circuitry}, goal-directed behavior \cite{miller2001integrative}, temporal organisation and reasoning \cite{fuster2001prefrontal}. \citet{romo1999neuronal} have shown that PFC neurons of non-human primate can maintain information about a stimulus for several seconds. Their firing rate was correlated with the coding of a specific dimension (frequency) of the stimulus maintained in memory. When Machens et al. \cite{machens2010functional} later re-analyzed the data of this experiment, they showed that the stimulus was actually encoded over a subpopulation using a distributed representation. Similarly, when~\citet{rigotti2013importance} analyzed single neuron activity recorded in the lateral PFC of monkeys performing complex cognitive tasks, they found several neurons displaying task-related activity. Once they discarded all the neurons that were displaying a task-related activity, they were still able to decode task information with a linear decoder and proposed that the PFC hosts high-dimensional linear and non-linear mixed-selectivity activity. Here, we can draw an interesting parallel between this linear-decoder analysis and the reservoir computer paradigm as originally proposed by \cite{buonomano1995temporal,dominey1995complex,jaeger_echo_2001,maass2002real}. In computational neuroscience, reservoirs are often used as models of generic neural circuits~\cite{hoerzer2012emergence,maass2002real,sussillo2014neural}. 
In particular, several authors used them to model cortical areas such as PFC~\cite{dominey1995complex,enel2016reservoir,hinaut2013real,mannella2015selection,hinaut2015corticostriatal}.

\citet{bechara1998dissociation} showed that working memory and decision making depend on separate anatomical PFC substrates. Correspondingly, \citet{dambre2012information} demonstrated the existence of a universal trade-off between the non-linearity of the computation and the short-term memory in the information processing capacity of any dynamical systems, including echo state networks \cite{jaeger_echo_2001}. In other words, the hyperparameters used to generate an optimal reservoir for solving a given memory task would not be optimal for a non-linear computation task. This conclusion is quite puzzling when one considers models such as long short-term memory (LSTM) networks \cite{gers_learning_2000, hochreiter_long_1997} that have been shown to solve complex tasks with long-term temporal dependencies\footnote{Long-term temporal dependencies imply non-linear computations if units have non-linear activation function.}. However, these models take advantage of an explicit gating mechanism inside each unit in order to store values for long periods of time\footnote{This does not actually refute the trade-off proposal by \cite{dambre2012information}.}.
The LSTM network uses explicitly engineered units that enables to store a value for long time spans by the use of gating mechanism. 
However, there is no reason to think that the brain has such engineered mechanisms, especially because the learning algorithm, back propagation through time (BPTT)~\cite{werbos1990backpropagation}, is unfolding time: this would be like if the brain could virtually duplicate it state values (for hundreds or more time points) in order to learn a particular time-dependency between two events for instance. 
With this study (and future ones) we want to explore how this gating mechanism could be performed without such engineered mechanisms, and explore how it could be performed with populations of neurons with less a priori constraints (e.g. random RNNs).

In the meantime, \citet{jaeger2012long} explored the capabilities of an ESN to maintain temporal information during long time spans and to exploit it for solving specific tasks.
He considered the suite of synthetic tasks originally proposed by \citet{hochreiter_long_1997} in order to test their LSTM model.
These synthetic tasks were also reused by \citet{martens2011learning} using an Hessian-free optimization of Recurrent neural networks (RNN).
\citet{jaeger2012long} showed that a correctly designed ESN could handle such tasks with the same success criteria originally proposed by \citet{hochreiter_long_1997}. 
However, the study was limited to transient short-term memory, and not on \enquote{attractor-like} working memory mechanisms. 
Conversely, other studies have focused on reservoirs with dedicated outputs acting as working memory (WM) units \cite{hoerzer2012emergence,pascanu_neurodynamical_2011}, which exploit different sub-regions of the reservoir high-dimensional space as \enquote{attractors}~\cite{sussillo_opening_2013}. 
These working memory units were trained to store binary values that were input-dependent; these WM units had feedback connections projecting to the reservoir. 
Thanks to these WM units, it enabled the reservoir to access and use such information, freeing the system to rely only on the reservoir short-term dynamics to maintain WM information.
\citet{pascanu_neurodynamical_2011} used up to six binary WM units to store information in order to solve a nested bracketing levels task.
With a Principal Component Analysis \citet{pascanu_neurodynamical_2011} showed that such binary WM units drive the reservoir in lower dimensional \enquote{attractors}\footnote{The WM units drive the reservoir to particular sub-regions of the high-dimensional space. These WM units act as \enquote{attractors} of the entire system, ie. if no particular input is seen, the system stays in the same sub-region.}.
Additionally, \citet{hoerzer2012emergence} showed that analog WM units (but encoding a binary information) actually drive the reservoir into a lower dimensional space (i.e. 99\% of the variability of the reservoir activities are explained by fewer principal components when there is feedback from WM units to the reservoir).
In this study, we want to extrapolate this idea of \enquote{attractor driven by WM units}  to WM units that have continuous values and not constrained to a population of binary values.

The paper is organized as follows: Section \ref{section:methods} introduces the different tasks and the reservoir design while Section \ref{section:results} introduced results on the different tasks. A discussion of the study is given in Section~\ref{section:discussion}.

\section{Methods and experiments}
\label{section:methods}

In order to better understand the properties of working memory mechanisms, we explore\footnote{\label{footnote:github}The code and the details on the hyperparameters are available at https://github.com/anthony-strock/ijcnn2018} the idea of a reservoir model connected to working memory (WM) units.
More specifically, we are interested in the capacity of a reservoir to store an arbitrary information (e.g. a real value instead of binary values like in~\cite{hoerzer2012emergence,pascanu_neurodynamical_2011}) from a streamed input by using a trained gating-like mechanism. 
As pictured in Table~\ref{table:summary-results}, we will explore six versions of a simple architecture composed of an Echo State Network (ESN) and optionally a WM unit and/or an output containing the result of a non-linear computation.
In a first task, we consider a model with two inputs: random values in [-1,1] changing over time and trigger events, signaling when to store information. The output should be clamped to the value given with the last trigger event: this forms a continuous WM unit that will change its value each time a new trigger is given. 
In a second task, we study the usefulness of such WM unit in order to solve a nonlinear task: multiplication of the input by the memorized value. 

\subsection{Model: Echo State Network}
\label{subsection:esn}

In this work we use an Echo State Network (ESN)~\cite{jaeger_echo_2001} with leaky neurons, and feedback from output to the reservoir described by the following update equations:

\begin{align}
  \label{eq:leak-update}
  x[n] &=  (1-\alpha) x[n-1] + \alpha \tilde{x}[n] \\
  \label{eq:reservoir-update}
  \tilde{x}[n] &= \tanh(W x[n-1] + W_{\text{in}} [1;u[n]] + W_{\text{fb}} y[n-1]) \\
  \label{eq:output-readout}
  y[n] &= f(W_{\text{out}} [1;x[n]])
\end{align}

where $u[n]$, $x[n]$ and $y[n]$ are respectively the streamed input vector, the vector containing the reservoir activations, and the output vector at time $n$. $W$, $W_{\text{in}}$, $W_{\text{fb}}$ and $W_{\text{out}}$ are respectively the recurrent, the input, the feedback and the output weight matrices. $[.;.]$ stands for the concatenation of two vectors, $\tanh$ (hyperbolic tangent function) and $f$ (linear or piece-wise linear, the identity function clamped to $-1$ in ]$-\infty$,-1] and to $1$ in [$1,+\infty$[)\footnote{$f$ is chosen piece-wise linear only for Table~\ref{table:rmse-results} and Figure~\ref{figure:drift-error}} are applied element-wise and $\alpha$ is the leaking rate.

The matrix $W$ is first randomly uniformly sampled between $-0.5$ and $0.5$ and then it is rescaled in order to set its maximal absolute eigenvalue, a.k.a spectral radius, to the chosen one. The matrices $W_{\text{in}}$ and $W_{\text{fb}}$ are both sampled uniformly between a value $s$ and its opposite $-s$. These two values are respectively called the input and the feedback scaling. In all the results, the reservoir contains 100 neurons.

As illustrated in the figures of Table~\ref{table:summary-results}, only the output weights will be learned; input, recurrent and feedback weight are generated randomly and kept fixed. The output weights are learned using ridge regression with teacher forcing~\cite{lukosevicius_practical_2012}, like stated in Equation~\ref{eq:learning}. When there are two output weights to train they are learned both at the same time considering the output vector to be 2-dimensional.

\begin{align}
   W_{\text{out}} &= Y X^T (X X^T + \beta I)^{-1}
   \label{eq:learning}
\end{align}

where $X$ is the concatenation of the reservoir activities at all time steps with a bias vector at 1, each row corresponding to a time step. Similarly, $Y$ is the concatenation of desired outputs and $\beta$ is the regularization parameter.

\subsection{Task 1: Storing a triggered real value} 


For this first task, the model used is the \enquote{Memory only} architecture (see Table~\ref{table:summary-results}).
An example of the inputs and outputs is shown in Figure~\ref{figure:example-model-memory-task}. 

\begin{figure}[!ht]
    \centering
    \includegraphics[width=\textwidth]{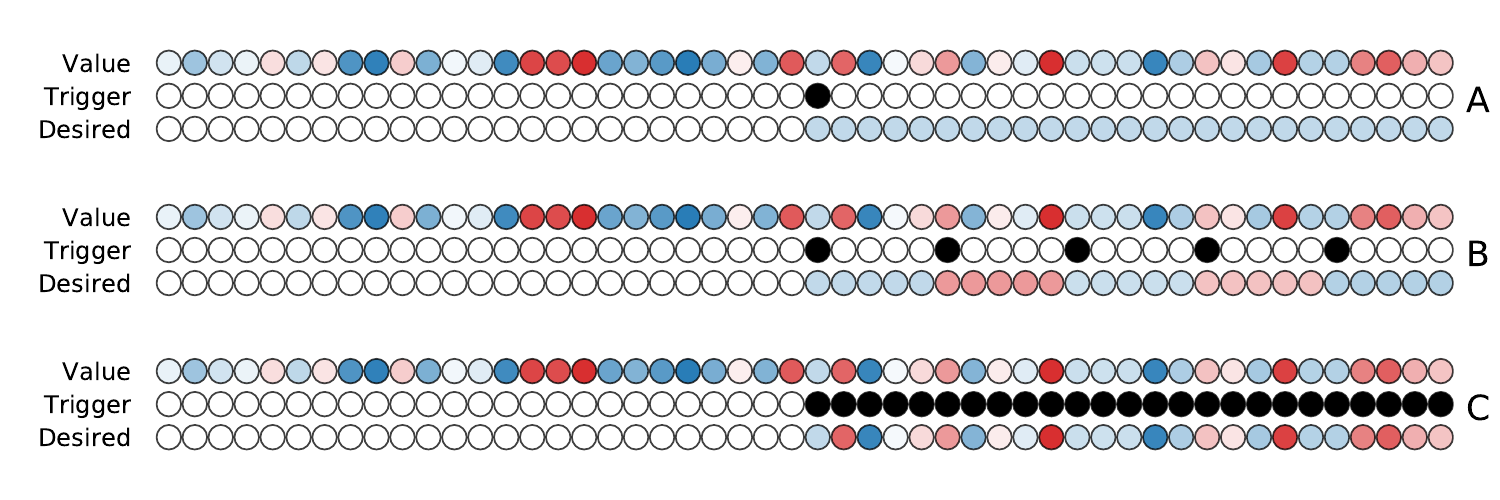}

    \caption{Three scenarios on which the trained models are tested. A. Single update scenario. B. Periodic update scenario. C. Continuous update scenario. Each column represents a time step. First two lines: input composed of the values and the trigger. Bottom line: desired output.
    }
    \label{figure:store-tasks}
\end{figure}

In this task, the model receives two different kinds of inputs: an input that indicates the value to be maintained when a trigger occurs, and an input that indicates the triggers. Before the first trigger event, the value to be stored is zero. As can be seen in Figure \ref{figure:store-tasks} three scenarios were studied.

\subsection{Task 2: Product of triggered value and current value}
\label{subsection}


Here we study a variant of the previous task where the output is a combination of the stored value and the current input signal.
The inputs are kept the same.
At each time step, the readout unit has to output the product of two values: the value of the current time step and the value synchronized with the last trigger (i.e. the stored/memorized value).
As before when no triggers occurred yet, the last value to be memorized should be zero, and so does the product.

The motivation to study this task is as follows. If the previously trigged value is available to the reservoir, the multiplication task is trivial (such a case is called {\em Oracle}).
However, if this previously triggered value is not easily available, the task becomes increasingly complicated as time runs, especially for a reservoir of only 100 units.
In order to explore and compare the capabilities in different architectures (six in total, see Table~\ref{table:summary-results}), we make a combination of three different scenarios:
\begin{itemize}
	\item No explicit memory: The model is trained only to perform this task without any further help.
    \item Trained explicit memory: The model is trained at the same time to solve the product task and the gating task.
	\item Oracle explicit memory: The model has to perform the product task and the triggered value is given as a supplementary input by an oracle. Only the capability to compute the product is tested in such case. 
\end{itemize}


\subsection{Training data}
\label{subsection:training-data}

The training data consists of 100 input sequences of 100 time steps that were randomly sampled in the following way:
\begin{itemize}
    \item \textbf{Value} is obtained by sampling a uniformly distributed value between -1 and 1;
    \item \textbf{Trigger} is obtained by sampling a value being either 0 or 1. When the trigger is 1, we call it a trigger event. In total 5 trigger events occur and they are uniformly sampled between time step 30 and time step 49.
\end{itemize}

For all experiments, we used the same training data that was sampled once when searching for good hyperparameters. For the principal component analysis, the training data consisted of 100 input sequences of 10,000 time steps similar to the previous one. The difference is that there is a constraint on the delay between two triggers. In Figure \ref{figure:memory-pc} there are exactly 200 time steps between triggers whereas in Figure \ref{figure:scaling-pc} there are between 100 and 200 time steps between triggers, the delay between triggers was sampled uniformly.

\subsection{Test data}
\label{subsection:test-data}

If not stated differently, the test sequences were generated with the same guidelines as the training data. They consist of 100 other input sequences of 100 time steps. 

For \textit{task 1}, we tested the trained model on four different test sets.
We first tested our model on data with similar statistics than the training set. 
We named it the "training-like scenario".
For \textit{task 2}, for all variants, we tested the trained model on the same "training-like scenario".
Moreover, in order to understand how the working memory of the model worked in \textit{task 1}, we tested it on three other scenarios for which the network was not trained, in order to test the abilities of the system to generalize over different statistical inputs. These scenarios can be seen in Figure~\ref{figure:store-tasks}. They are particularly challenging in the sense that the model has not been trained with so long inputs. Each scenario was built using 100 input sequences of 10,000 time steps generated in the following way:
\begin{itemize}
    \item \textbf{Value} is obtained, as before, by sampling a uniformly randomly distributed value between -1 and 1;
    \item \textbf{Trigger} is obtained, unlike before, by choosing by hand when the trigger occurs. For the single update scenario (A) the trigger occurs only at time step 100. For the periodic update scenario (B) the first trigger occurs at time step 100 and then a trigger occurs every 1000 time step. Finally, for the continuous update scenario (C) the first trigger occurs at time step 100 and triggers will occur at every time step after that.
\end{itemize}

In order to generate an understandable view of how the task is behaving, we created two extra scenarios for the Figures~\ref{figure:example-model-memory-task}, \ref{figure:example-model-product-task}. For both, we have smoothed the inputs by convolving the random signal by an exponential window of width 10 and decay 5. For \textit{task 1} the inputs are input sequences of 100 time steps whereas for \textit{task 2}, in order to understand how the memory was used in the Trained explicit memory architecture, the inputs are input sequences of 200 time steps. All the other results are given for random inputs values without any smoothing.

\subsection{Evaluation procedure}
We took two different evaluation procedures. For both tasks, while using the training-like scenario we searched for best hyper-parameters in order to solve the task; we used the \textit{hyperopt} Python toolbox~\cite{bergstra2013hyperopt}.

First, for all architectures, in order to be able not only to find better hyper-parameters but to extract information on the influence of hyper-parameters on the task we decided to use a random search for 1000 hyper-parameters, the first 200 ones being chosen purely randomly and the following one using a Bayesian approach called Tree-structured Parzen Estimator (TPE)\footnote{We sampled from a very large range of values: we sampled log-uniformly the spectral radius (between 1e-10 and 1e10), the leaking rate (1e-10, 1.0), the input (resp. feedback) scaling (1e-10, 1e10) and the regularization parameter (1e-15, 1e-1).}. In order to evaluate the performance of the ESN, in this case, we generated 10 instances of the model using the sampled hyper-parameters. We trained all these ESNs on the same training data (see subsection~\ref{subsection:training-data}).

Secondly, in the three other scenarios on \textit{task 1}, we picked the best hyper-parameters found for this task\footref{footnote:github}. This time in order to evaluate the performance of the ESN on each scenario we generated 100 instances of the model. Once trained, we tested all instances on the 3 last scenarios (see subsection~\ref{subsection:test-data}). The following is performed the same way than in the training-like scenario.

Before feeding a new input sequence to a reservoir we initialized all its activations to zero. The error we consider at a given time step is the absolute difference between the produced output and the desired output. The global performance criterion we consider is the Root Mean Square Error (RMSE).

\section{Results}
\label{section:results}

\subsection{Task 1: Storing a triggered real value}

\begin{table}
    \begin{tabular}{|c|c|}
        \hline
        Scenario & RMSE (Mean$\pm$Std) \\ 
        \hline
        Training-like & 2.28e-4$\pm$1.90e-4 \\
        Single (A) & 4.55e-2$\pm$4.19e-2 \\
        Periodic (B) & 5.04e-3$\pm$4.17e-3 \\
        Continuous (C) & 9.25e-5$\pm$3.66e-5 \\
        \hline
    \end{tabular}
    \caption{Global performances for the four scenarios for task~1: Mean, standard deviation of the Root Mean Square Error (RMSE).}
    \label{table:rmse-results}
\end{table}
\begin{figure}
	\centering
	\includegraphics[width=\textwidth]{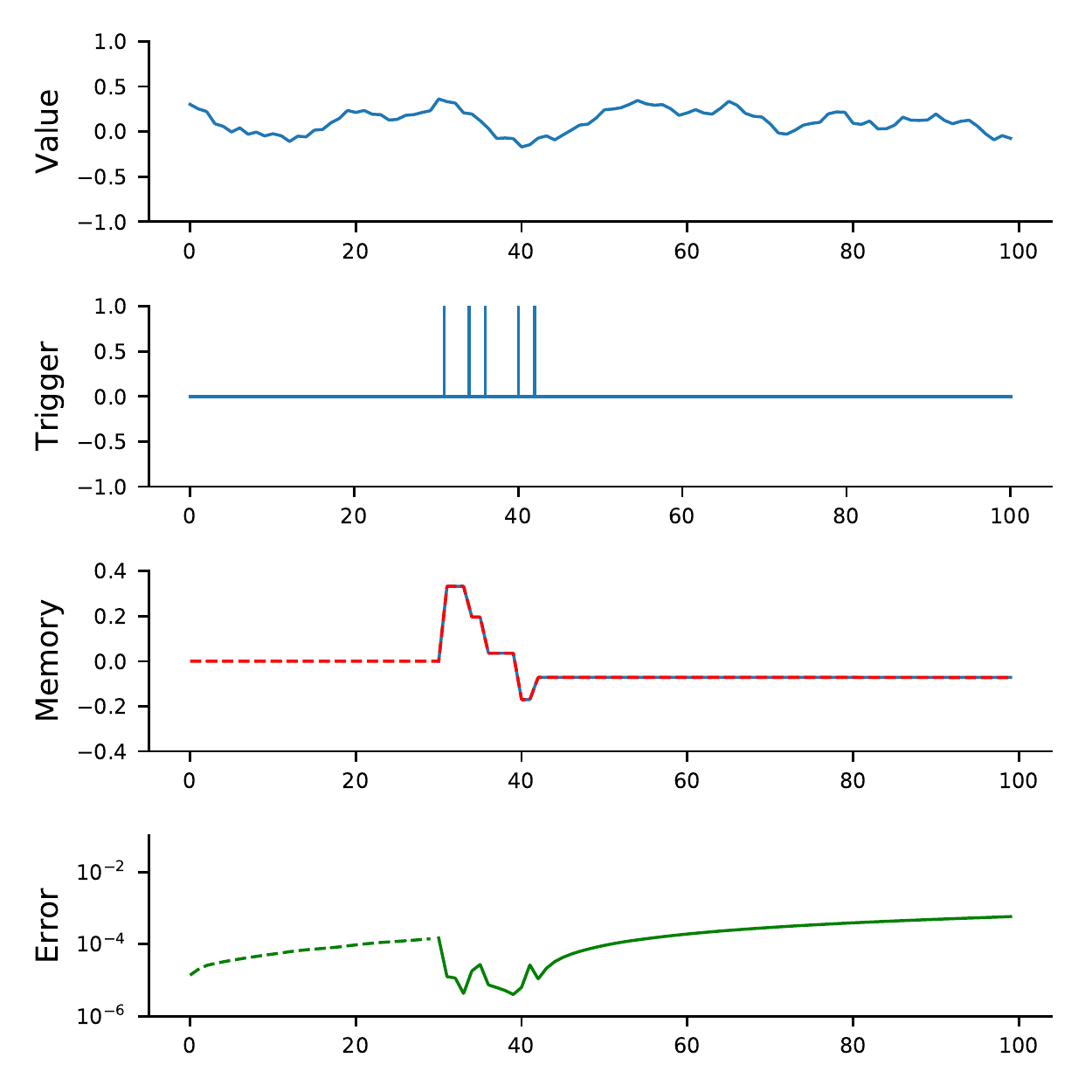}
    \caption{Behavior of the model trained to perform \textit{task 1} (gating task) in a visually understandable case. The model is trained with the uniform random inputs as values and tested on a smoothed version of a uniform random signal.
\textbf{Value}: values received by the model, in this example they are obtained by filtering with an exponential window of width 10 and decay 5 a uniform random sample between -1 and 1.
\textbf{Trigger}: triggers received by the model obtained in the same way than in the training scenario.
\textbf{Memory}: the desired output is in blue which starts after the warm-up, and the produced output by the model in dotted red. 
\textbf{Error}: the absolute difference between the desired and produced output in log scale.
Inputs shown are different from what the network has been trained on: they are pseudo-random instead of uniformly random in order to give a more comprehensible visualization.}
	\label{figure:example-model-memory-task}
\end{figure}
\begin{figure}
    \centering
    \includegraphics[width=\textwidth]{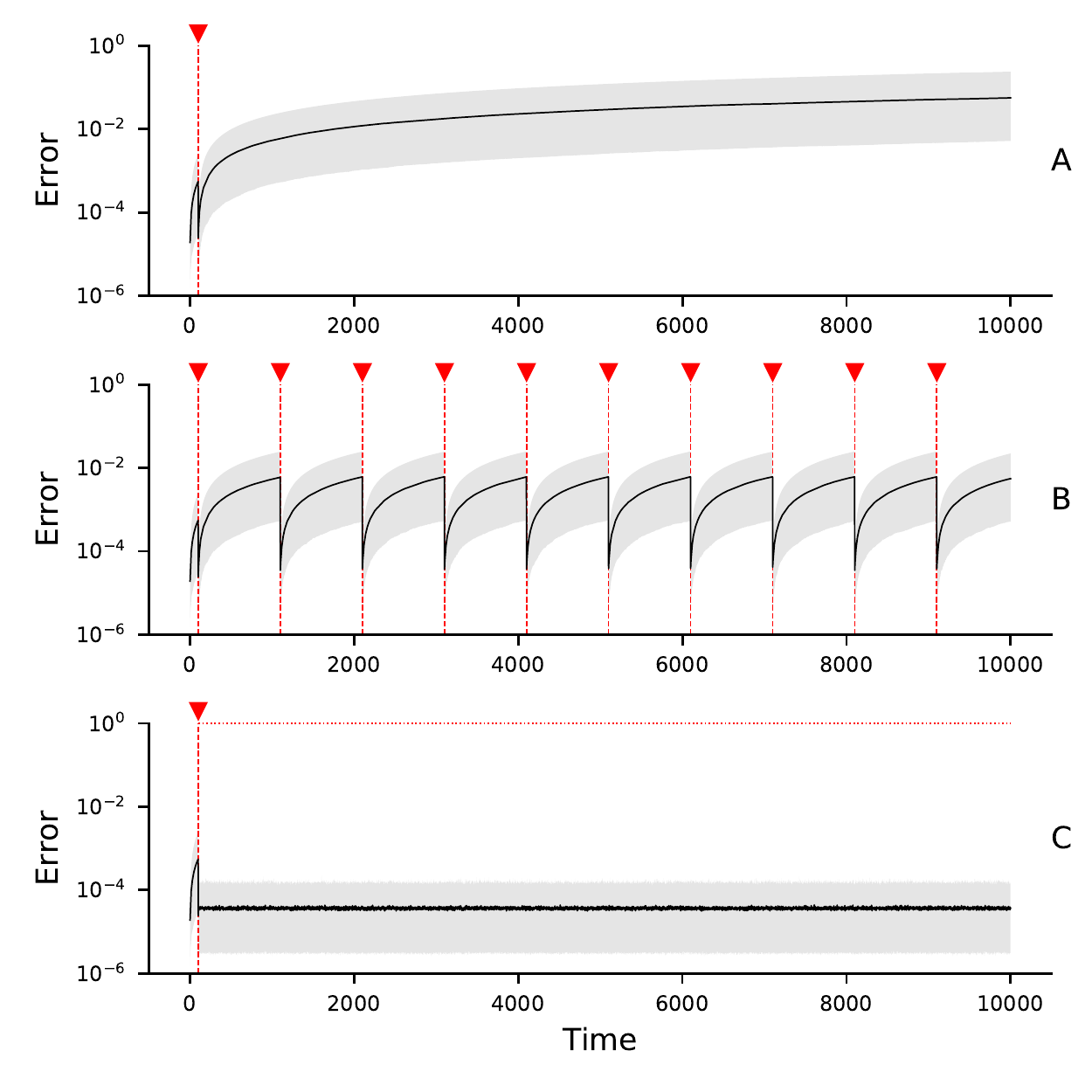}
    \caption{Statistics on the evolution of the error over time for three testing scenarios (task 1). Median (black line), 5th percentile (bottom edge of the gray surface) and 95th percentile (top edge) of the absolute error varying over time.
    A. Single update scenario. B. Periodic update scenario. C. Continuous update scenario.
    The red vertical lines represent the triggers, the constant update being shown by a horizontal red dot line.}
    \label{figure:drift-error}
\end{figure}
\begin{figure*}
    \centering
    \includegraphics[width=.4\textwidth]{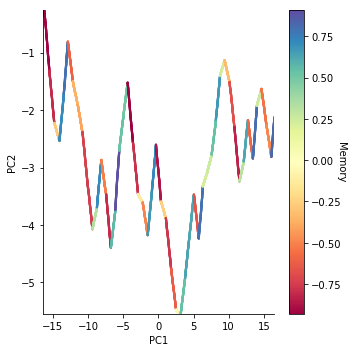}
    \includegraphics[width=.4\textwidth]{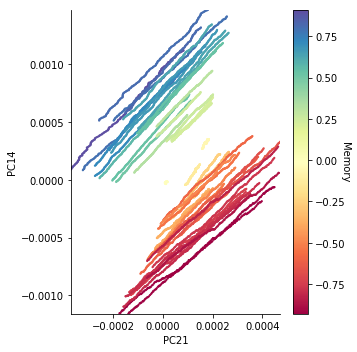}
    \caption{Task 1, \textit{Memory only} architecture. {\bf Left} Evolution of the value memorized in the working memory unit along the two principal components that explain more than 99\% of the variance of the internal activities. 
    First PC is correlated with time: we can see the sequence of the memorized values.
    {\bf Right} Evolution of the value memorized in the working memory unit along the principal components that are the most correlated with
    the memorized value.}
    \label{figure:memory-pc}
\end{figure*}

First, we studied the global performance of the model relatively to the task.
Table~\ref{table:rmse-results} reports a good generalization on the training-like scenario with a mean RMSE of 1.63e-4 ($\pm$1.14e-4) and the generalization remains good for all the other scenarios.
In all cases, the RMSE is equal or smaller than 5e-2.
The single update scenario (A) (RMSE about 5e-2) is harder for our model than the periodic update scenario (B) (RMSE around 5e-3) which,
in turn,
is harder for our model than the continuous update scenario (C) (RMSE around 1e-4).

Then we analyzed how was evolving the error over time. In Figure~\ref{figure:example-model-memory-task} we show an example of the evolution of the output. We can note two properties on this figure: the error remains stable around 1e-4 during the whole run and it even goes down to 1e-4 during warmup way before it has been trained to be. The latter might suggest that a smaller warmup could be used.
A very light drift seems also to appear: we created the scenarios A, B, and C in order to study more extensively this drift. In the following, we focus on the variation of the error over time shown in Figure~\ref{figure:drift-error}.
In all cases, after 10,000 time steps, the error is far below 1e-1.
Moreover,
one can see that after a trigger event,
the error gets around 1e-4,
and in 1,000 steps the error has not even the time to reach 5e-2.
Furthermore, in scenario B, the mean error 10 time steps after a trigger is 1.51e-04 ($\pm$7.87e-04), and for 100 time steps, it is 8.88e-04 ($\pm$1.14e-03).

To summarize, we can notice a constant increasing drift of the output when there is no trigger~(A).
However, this drift can be bounded thanks to a periodic trigger that feed a new value into the reservoir~(B), the faster the update the lower the bound~(C).
For tasks for which the gated value has to be kept for a very long amount of time with high precision (and if one do not want to increase the reservoir size of 100 units), a simple counting mechanism could provide this periodic trigger. This could probably be implemented with the same reservoir.

In order to understand how 
the memorized value
is encoded in the reservoir, we performed a principal component analysis (PCA) on the reservoir states. 
Similarly to \citet{pascanu_neurodynamical_2011} and \citet{hoerzer2012emergence}, this PCA analysis shows that the dynamic evolves in a low dimensional space where two components explain more than 99\% of the variance.
In fact, even the first component alone explains more than 89\% of the variance.
However, 
the first two components do not seem directly related to the stored value in the WM unit. 
Figure~\ref{figure:memory-pc} (Left) shows the evolution of the reservoir for a specific input sequence. In this figure, we can see that the first component is correlated with the evolution of time as it keeps increasing with time, the correlation with time is actually near 1. The second component doesn't contain directly 
the memorized value
but it's discrete derivative\footnote{We call discrete derivative of a value that evolves in a discrete time, the evolution of the difference between the value at a time and at the previous time.} does: the correlation between its discrete derivative and
the memorized value
is near 1. Combining the two, we can see that on the first two components the dynamic evolves piece-wise linearly and 
the memorized value
is actually the slope of the lines.
In order to find components that could explain better how is stored the memorized value, we also watched at the components that were the most correlated with
the memorized value.
Not so surprisingly that was not at all the one of before as
the memorized value
is not linearly encoded in any of them. These components explain not much about the variance but, as we can see in Figure \ref{figure:memory-pc} (Right), the same kind of phenomena appears. When a given value is stored, these components evolve on a line, but now that's not the slope that contains
the memorized value
all the slope are similar, that's the offset of the line. Moreover, we can note that in these components the values stored in memory are better distributed in the space. When a new value is stored these components jumps in another place. One can notice a linear color gradient along the axis perpendicular to the lines. So, similarly than in \cite{machens2010functional} for neural recording, the components that explain the best the variance might not be the best ones to explain how
the memorized value
is encoded.

\subsection{Using a working memory unit to solve a complex task}

\begin{table}
	\centering
	\begin{tabular}{|>{\centering\arraybackslash}m{1.3cm}|>{\centering\arraybackslash}m{5cm}|>{\centering\arraybackslash}m{1.5cm}|}
    	\hline
    	\g{Task}& \g{Architecture}&\g{RMSE}\\
       	\hline
		Memory only&\includegraphics[width=5cm]{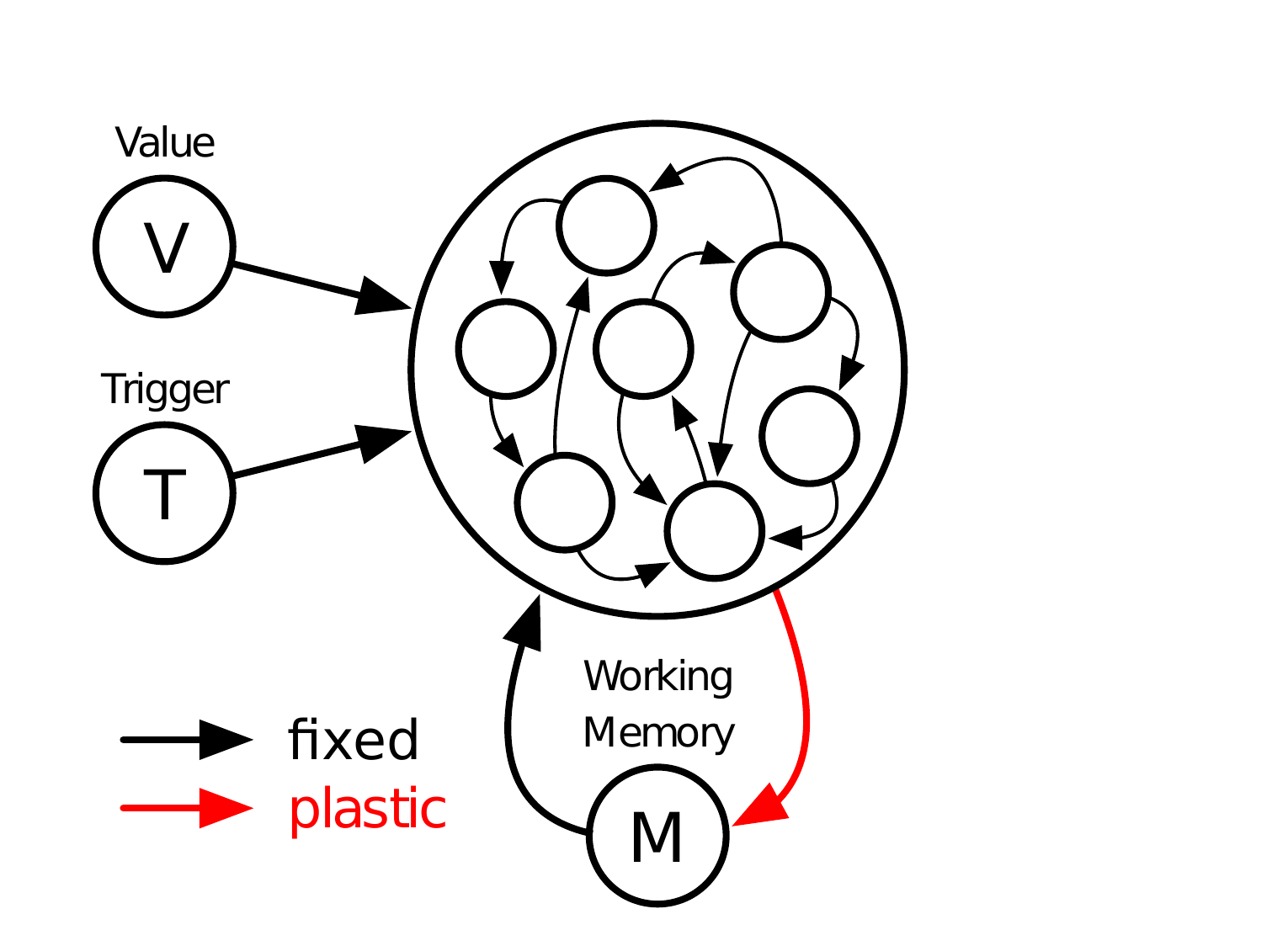}& \makecell{\textbf{1.55e-4}\\$\pm$7.42e-5}\\
        \hline
		No explicit memory&\includegraphics[width=5cm]{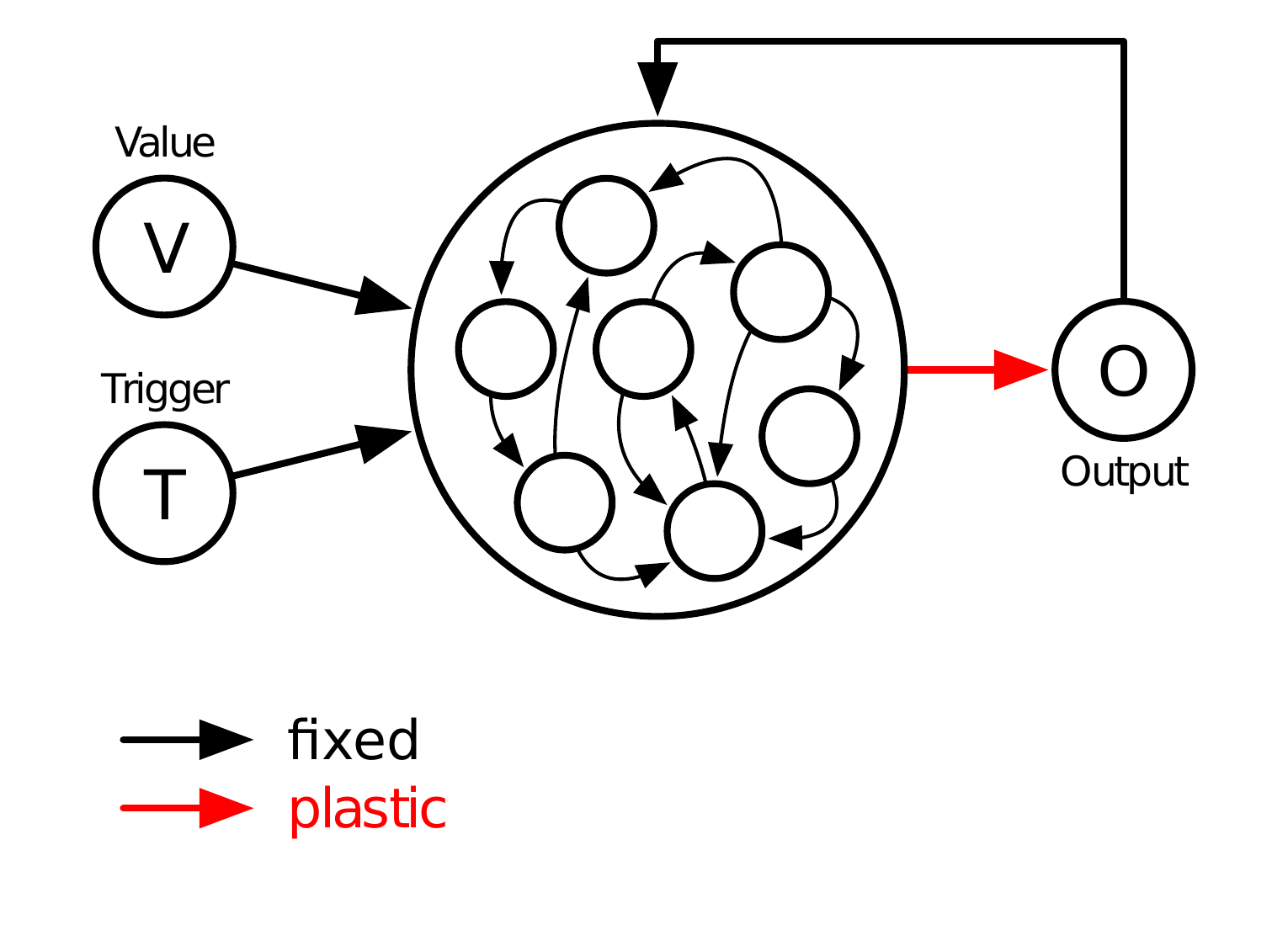} &\makecell{3.03e-1\\$\pm$4.53e-4}\\
        \hline
		\makecell{No\\explicit\\memory\\(No FB)}&\includegraphics[width=5cm]{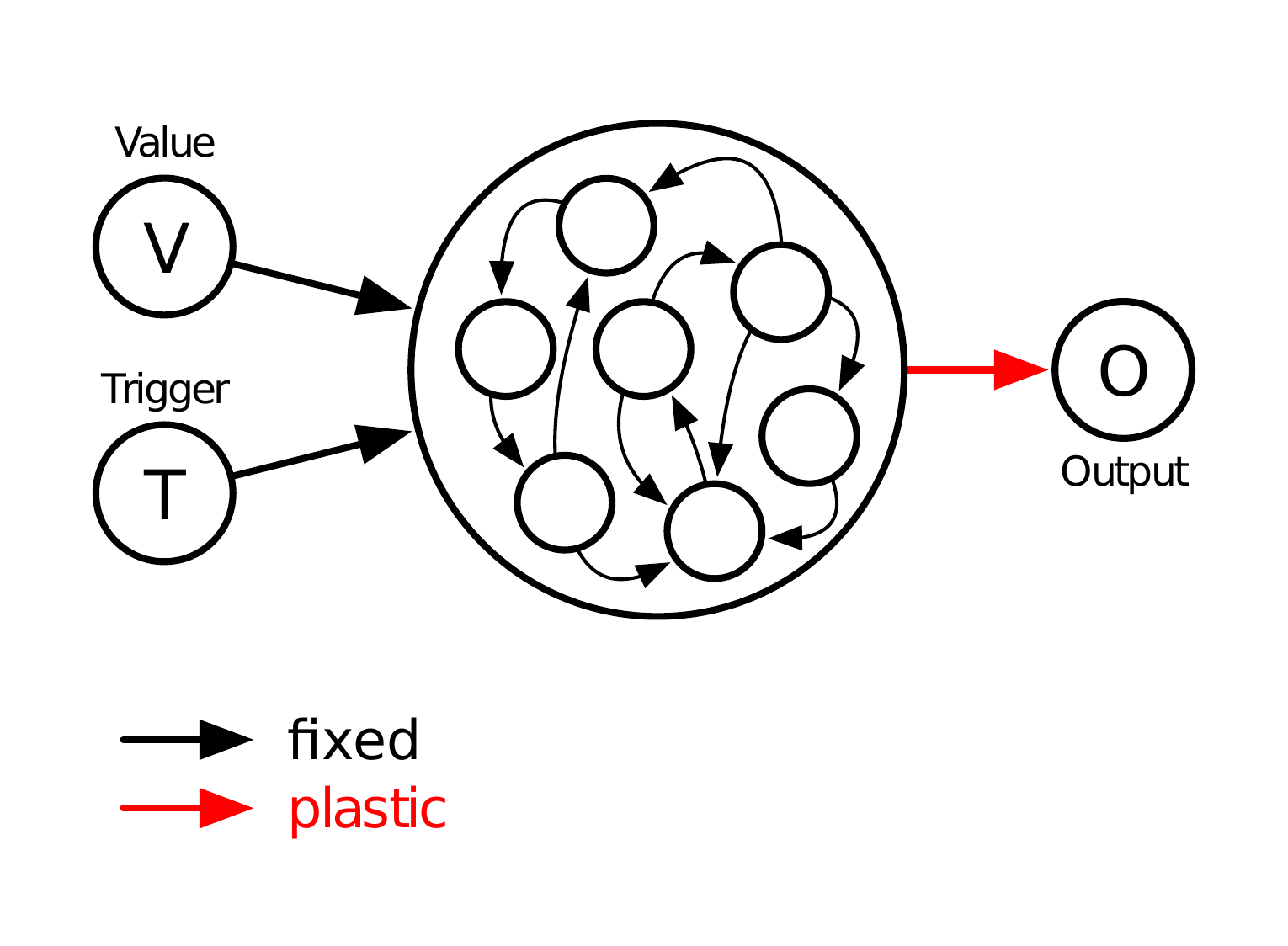} & \makecell{3.05e-1\\$\pm$3.67e-4}\\
        \hline
		Trained explicit memory&\includegraphics[width=5cm]{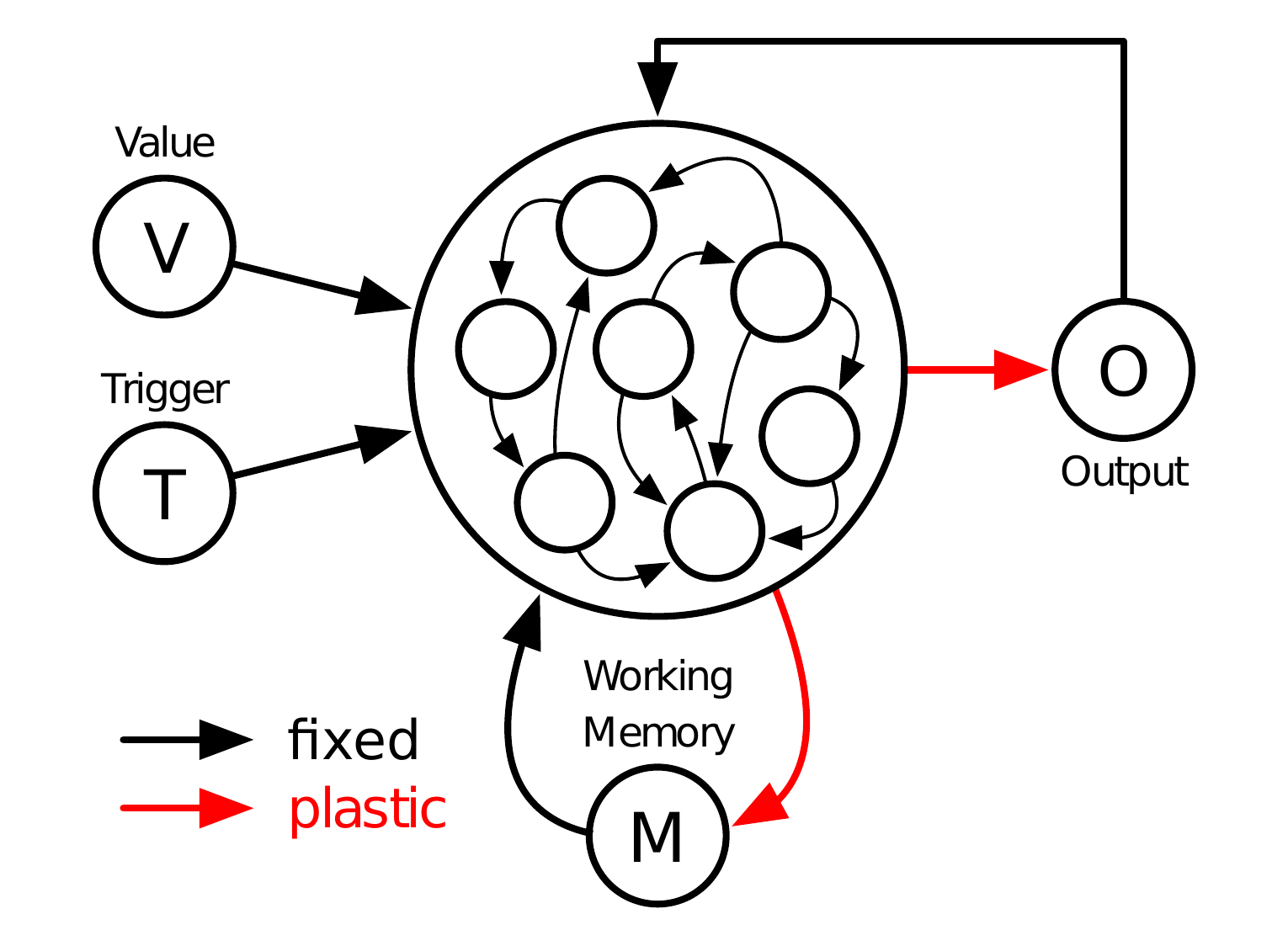}&\makecell{\textbf{7.26e-4}\\$\pm$1.88e-4}\\
        \hline
		Oracle explicit memory&\includegraphics[width=5cm]{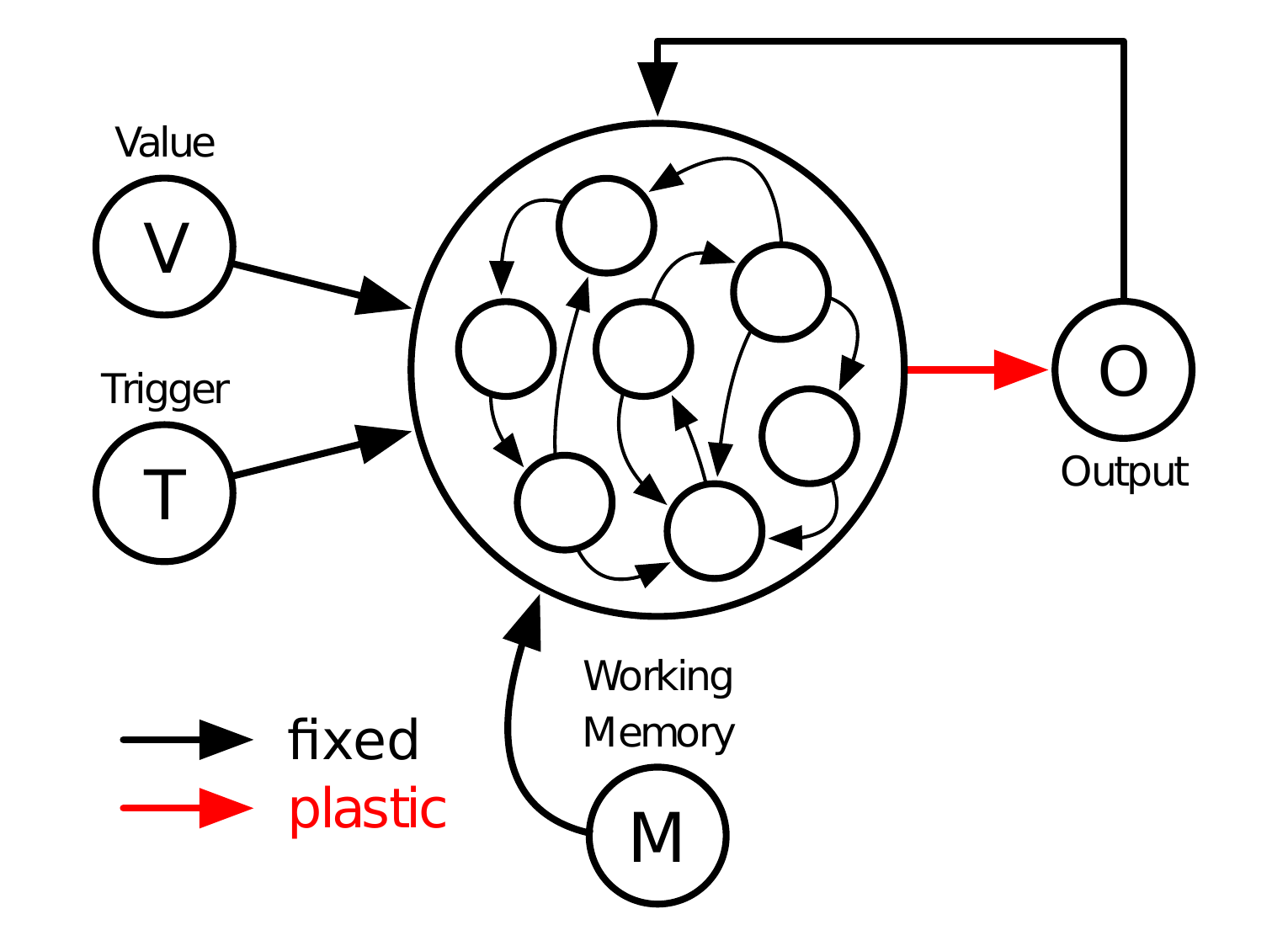} &\makecell{1.99e-4\\$\pm$3.15e-5}\\
        \hline
		\makecell{Oracle\\explicit\\memory\\(No FB)}&\includegraphics[width=5cm]{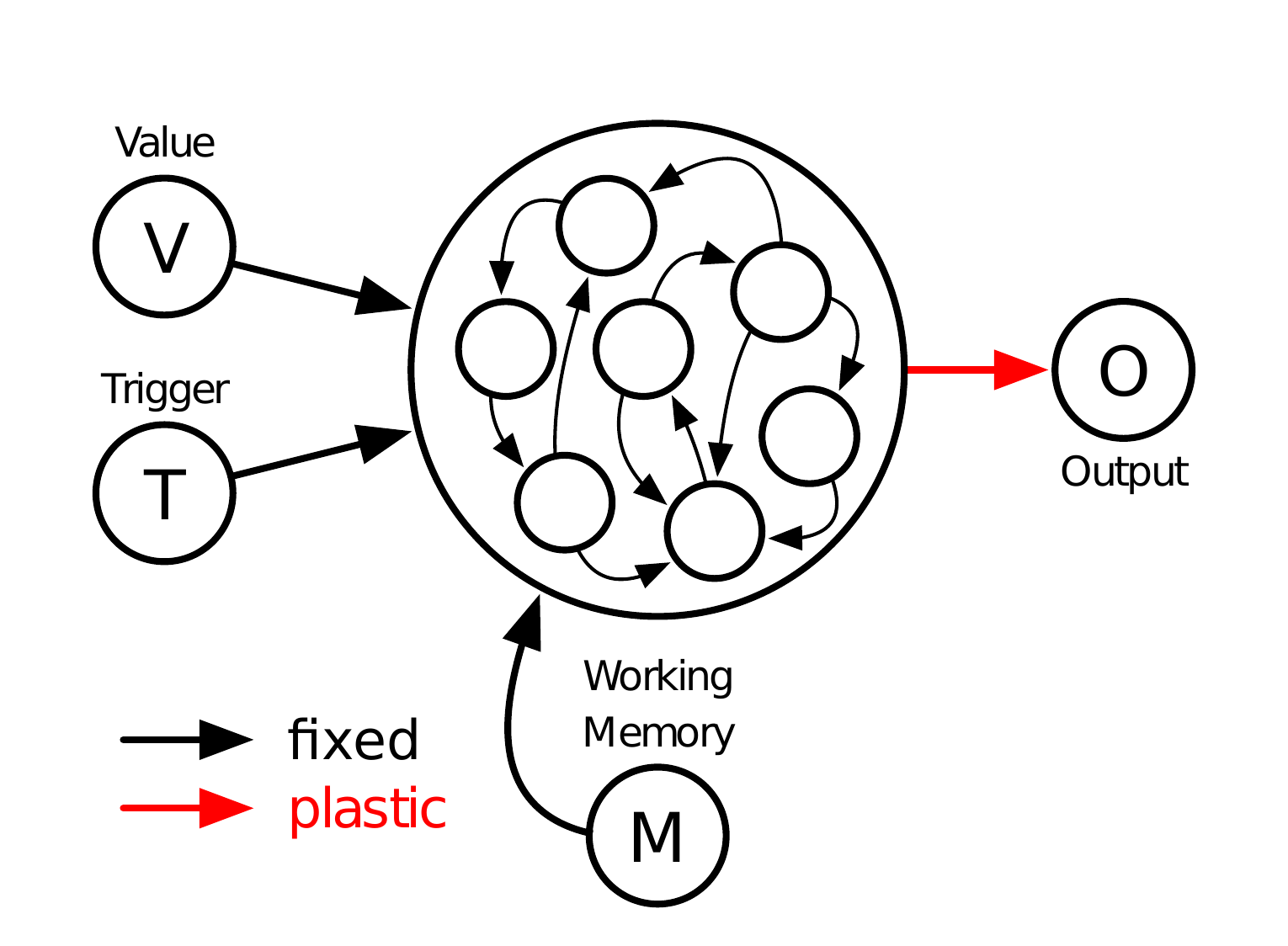} &\makecell{7.10e-5\\$\pm$2.65e-5}\\
        \hline
    \end{tabular}
    \caption{Summary of the performance for the best hyper-parameters found by Bayesian optimization. The number of neurons was fixed to 100.}
    \label{table:summary-results}
\end{table}
\begin{figure}
	\centering
	\includegraphics[width=\textwidth]{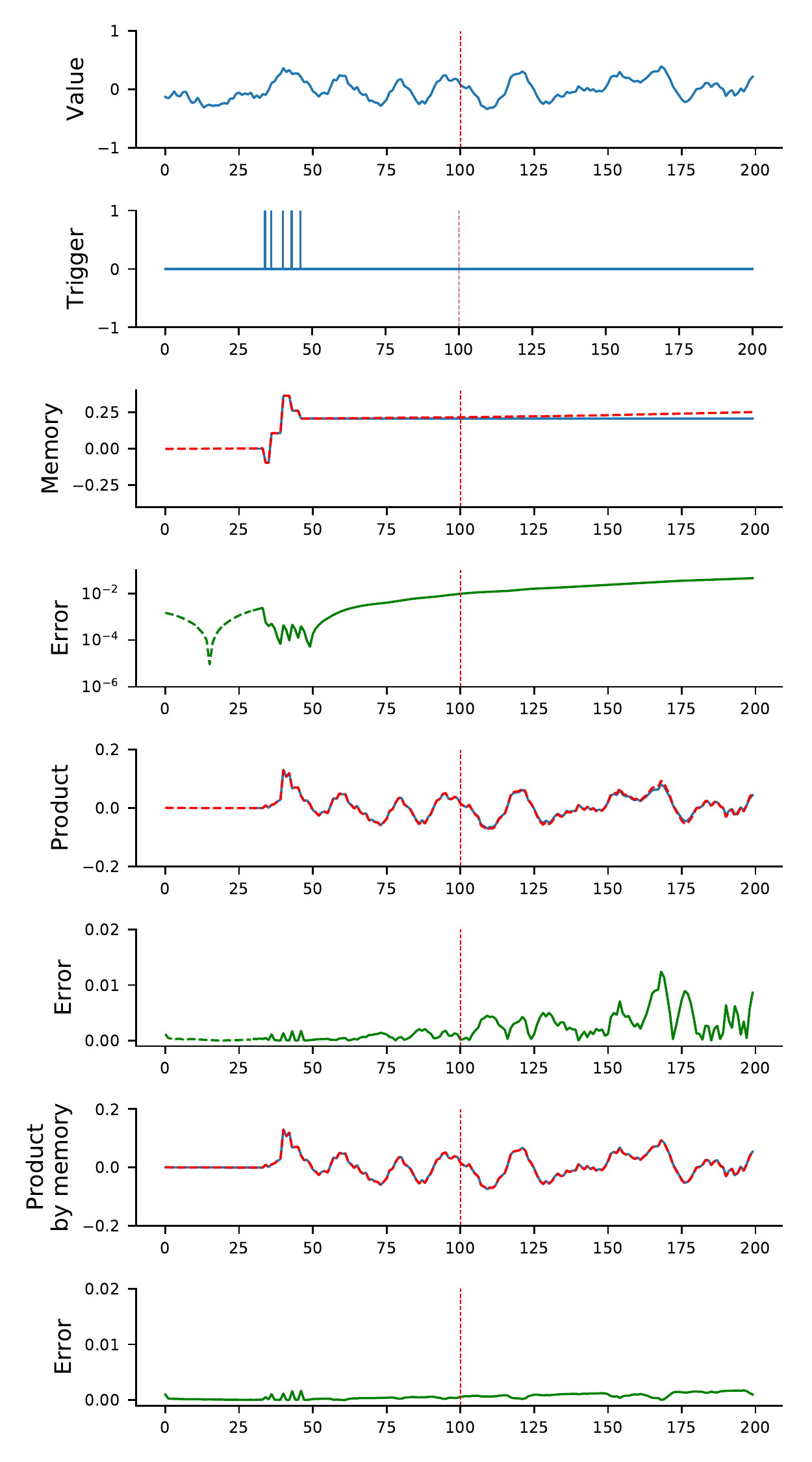}
    \caption{Illustration of the behavior of the model trained to perform the product task in a more understandable case. \textbf{Value}: values received by the model. 
\textbf{Trigger}: triggers received by the model. \textbf{Memory}: two curves superposed, the desired (working) memory output in blue which starts after the warm-up, and in dotted red the memory output obtained. \textbf{Product}: two curves superposed, in blue the desired product output (i.e. the product of the desired memory and the value received), and in dotted red the product output obtained.
\textbf{Product by memory}: two curves superposed, in blue the desired product output if it had to be consistent with the memory stored (i.e. the product of the actual memory and the value received) and in red dotted the product output obtained. \textbf{Error}: the absolute difference between the desired and produced output (in log scale for Memory). Each plot corresponds to the error of the plot just above. The desired output and the error are shown after the warm-up that has been used during training. The vertical red dot line is here to represent the time step after which the model has not been trained to perform the task.
(Inputs shown are different from what the network has been trained on.)
} 	
    \label{figure:example-model-product-task}
\end{figure}
\begin{figure*}
    \centering
    \includegraphics[width=.4\textwidth]{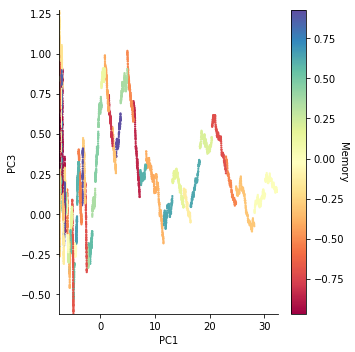}
    \includegraphics[width=.4\textwidth]{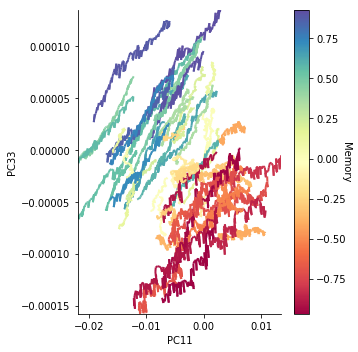}
    \caption{Task 2, \textit{Trained explicit memory} architecture {\bf Left} Evolution of the value memorized in the working memory unit along
    the first and the third principal components (PC).
    The first PC is correlated with time: we can see the sequence of the memorized values.
    {\bf Right} Evolution of the value memorized in the working memory unit along the PCs of the internal activities that are the most correlated with the memorized value. In order to better see the temporal evolution, the dots at a time step is connected to the one at the previous time step, except in the case where the model receives a trigger (in order to avoid meaningless artifacts).
    }
    \label{figure:scaling-pc}
\end{figure*}

Given the results from task 1, we know that we have a unit able to store a real value.
In the following, we studied how such ability can help to perform another task: output the product between the current input and a previous input indicated by a trigger.

First, we tried to answer 3 questions and the results we used to provide answers are summarized in Table~\ref{table:summary-results}:
\begin{itemize}
\item \textit{Does the presence of an explicit memory help?} The mean RMSE obtained without explicit memory is at least 3 order of magnitude above the ones using a working memory unit. Moreover, the mean RMSE in the latter case is at 7.26e-4~$\pm$1.88e-4, so the task is actually performed quite well. Thus, having a working memory unit greatly helps.
\item \textit{How does the training between memory and product interfere ?} To answer this question, we replaced the WM unit with an Oracle.
The difference between the architecture with the WM unit compared with the Oracle is lower than one order of magnitude: 7.26e-4~$\pm$1.88e-4 against 1.99e-4~$\pm$3.15e-5 for the Oracle. Therefore, the fact of training the working memory at the same time as the product output does not seem to interfere much with the performance obtained on the product.
\item \textit{Is the feedback helping when there is no WM unit?} In the two cases studied, the fact of having or not a feedback did not change qualitatively the results.
\end{itemize}

Afterward, we studied the behavior of the model. Here, we focus on the "Trained explicit memory" architecture:
with feedback from both the output and the working memory unit.
In Figure~\ref{figure:example-model-product-task}, we show how the model behaves with simplified filtered random input values.
On the first 100 time steps, the model can keep a working memory unit stable as it has been trained to and perform the product with a very small error.
By comparing the error in the curves below the "Product" and "Product by memory" curves,
we can see that the value used to perform the product is the actual stored value and not the theoretical value (that should have been stored).
The model seems to use the working memory unit as we would expect it to do.
This is similar to what \cite{hoerzer2012emergence} have shown, but with real values instead of binary values (or switches).

Finally, we studied the evolution of the internal activities of our model: as before, to reduce the dimensionality a principal component analysis (PCA) has been performed on the reservoir states.
Again as in \cite{hoerzer2012emergence}, the two first principal components contain a lot of information: here they can explain 99.8\% of the variance of the activations. However, in this case, the first component already explains more than 96\% of the variance (compared to 89\% for \textit{task 1}), and thus explains even more variability than earlier. As before, the first component is linked to time, but here the correlation is non-linear. However, the link between the memorized value and the second component is less clear than before. As shown in Figure \ref{figure:scaling-pc} (Left), with the third component we can see something similar (even if noisier) to the second component of the previous PCA analysis (Figure~\ref{figure:memory-pc}, Left). On the first and third components, the activity nearly evolves as lines and the slope of this line depends on the value stored: when the memorized value is negative (red, orange) it decreases, when it is positive it increases (green, blue). In this case, we also watched the two components that were the most correlated with the memorized value. As before, the activity seems to evolve noisily on parallel lines depending on the value stored, the blue (resp. red) lines, corresponding to positive (resp. negative) memorized values. 


\section{Discussion}
\label{section:discussion}

We have shown how a small group of randomly connected units is able to maintain an arbitrary value at an arbitrary time from a streamed input.
It is to be noted that the model has not been trained at memorizing (by heart) every possible value since there is virtually an infinite number of values between -1 and +1. What the model has actually learned is to gate an input value into a placeholder, a.k.a. a working memory unit.
This property of the model can be considered as a gated memory: a value enters the memory at the moment of the (input) trigger and is kept constant in face of incoming distractors (the continuous streamed input).
Such robustness is actually characteristic of a gated working memory: information enters while the gate is opened and is kept constant once the gate is closed.
In that regard, it is to be noted that previous works have addressed the question but in different ways. \citet{Lim:2013} have shown that a balanced cortical microcircuitry can give account of the maintenance of information. The model uses a corrective negative feedback that makes it robust against common perturbation but requires a balanced amount of excitation and inhibition. Similarly, \citet{Stern:2014} explains how random neural networks can exhibit a bi-stable behavior using strong local connectivity and random inter-unit connections.
In another kind of recent work 
Jaeger~\cite{jaeger2014controlling} use \textit{conceptors} to project and drive the dynamics in a sub-space of lower dimension.

We have also demonstrated how such explicit working memory is critical in solving a multiplication task.
Probably the most interesting point in this task is the following: the same model deprived of the explicit working memory fails at solving the task and exhibits very bad performances (error of 3.03e-1 instead of 7.26e-4).
This demonstrates the criticality of the presence of the working memory unit. 
Interestingly, these results are coherent with the \citet{dambre2012information} trade-off hypothesis: the \enquote{No explicit memory} model performs worse than the models with WM units because its dynamics are not able to perform the non-linear computations and the long-term memorization at the same time.

In this study, we did not try to find the optimal number of reservoir units needed for each task. Conversely, we voluntary limited the size of the reservoir to 100 neurons in order to see if such rather small reservoirs were sufficiently competitive. 
Moreover, even though we have optimized the hyperparameters of the model in order to find the best performances, the random nature of the network suggests that such working memory property is an intrinsic property of any (recurrent) group of neurons under some conditions (size, spectral radius, leak rate)\footnote{We choose some particular hyperparameters among many good enough ones: each parameter seems robust in a quite substantial range of values.}. Furthermore, even though it was not the main goal here, this study is a first step in exploring the plausibility of having a network of such small gated reservoirs being interconnected together, similarly as LSTM cells in an LSTM network.

\bibliographystyle{plainnat}
\bibliography{references}

\ifCLASSOPTIONcaptionsoff
  \newpage
\fi

\end{document}